\documentstyle[epsf,floats,prd,aps]{revtex}

\def\be{\begin{equation}}
\def\ee{\end{equation}}
\def\bea{\begin{eqnarray}}
\def\eea{\end{eqnarray}}
\def\bml{\begin{mathletters}}
\def\blea{\begin{mathletters}\begin{eqnarray}}
\def\elea{\end{eqnarray}\end{mathletters}}

\begin{document}
\draft
\wideabs{
\title{The Ori-Soen time machine}

\author{Ken D.\ Olum\footnote{Email address: {\tt kdo@alum.mit.edu}}}

\address{Institute of Cosmology \\
Department of Physics and Astronomy \\
Tufts University \\
Medford, MA 02155}

\date{June 1999; Revised December 1999}

\maketitle

\begin{abstract}
Ori and Soen have proposed a spacetime which has closed causal curves
on the boundary of a region of normal causality, all within a region where
the weak energy condition (positive energy density) is satisfied.  I
analyze the causal structure of this spacetime in some simplified
models, show that the Cauchy horizon is compactly generated, and argue
that any attempt to build such a spacetime with normal matter might
lead to singular behavior where the causality violation would
otherwise take place.
\end{abstract}

}
\def\thefootnote{\fnsymbol{footnote}}
\footnotetext[1]{Email address: {\tt kdo@alum.mit.edu}}
\def\thefootnote{\arabic{footnote}}

\narrowtext

\section{Introduction}
In the absence of any restrictions on the stress-energy tensor,
general relativity permits an arbitrary spacetime.  One simply writes
down the desired metric, computes the curvature, and solves Einstein's
equations in reverse to find the required matter content.  In
particular, the spacetime may contain closed timelike curves (CTC's),
future-directed timelike paths which return to the same point in
spacetime.  Thus general relativity always permits time travel unless
one restricts the matter content that one can use as a source.

To prove theorems about the properties of a spacetime, one uses energy
conditions, i.e., restrictions on the stress-energy tensor
$T_{\mu\nu}$.  Perhaps the most important of these is the weak energy
condition (WEC), which states that every timelike observer must see a
nonnegative energy density, i.e., that $T_{\mu\nu} V^\mu V^\nu\ge 0$
for any timelike vector $V^\mu$.  Tipler \cite{tip76,tip77} and
Hawking \cite{cpc} have shown that if a spacetime obeys the weak
energy condition, closed timelike curves cannot be produced in a
compact region.  They proved these theorems by considering the Cauchy
horizon, which is the boundary of the region whose entire past
intersects some initial surface $S$.  The Cauchy horizon is composed
of null geodesic generators which can have no past endpoints
\cite{Ha&El}.  If this horizon is to arise from a compact region, each
generator must wind around indefinitely inside this region as one goes
into the past.  This in turn requires the generators to originate as a closed
null geodesic called the fountain.  This means they must be defocused,
which is impossible if the weak energy condition is obeyed.

\begin{figure}
\begin{center}
\leavevmode\epsfbox{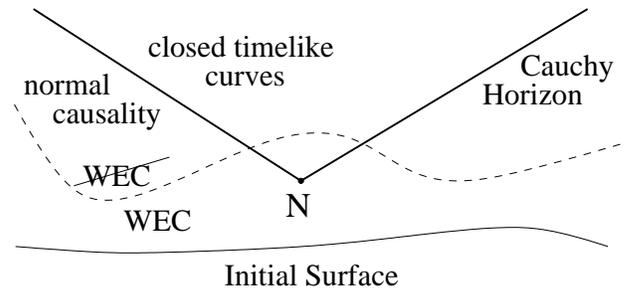}
\end{center}
\caption{The Ori-Soen spacetime.  The weak energy condition is obeyed
below the dashed line.  The thick line is the Cauchy horizon, which
separates the region with normal causality from the region with closed
timelike curves.  The point $N$ is a closed null geodesic.}
\label{fig:intro}
\end{figure}

However, Ori and Soen \cite{Ori1,Ori2,Ori3} exhibit a spacetime which
casts some doubt on the effectiveness of these theorems at preventing
the construction of time machines.  This spacetime contains a closed
null geodesic $N$; all points in the future of $N$ have causality
violation, while all points in its chronological past have normal
causality.  The weak energy condition is obeyed in a region of finite
size surrounding $N$.  The Ori-Soen spacetime is sketched in Fig.\
\ref{fig:intro}.

How can this spacetime evade the theorems of Tipler and Hawking?  The
answer is that $N$ is not in fact the fountain where the causality
violating region originates, but rather a place where the Cauchy
horizon terminates.  The origin of the causality violation lies
outside the region where the weak energy condition is obeyed.

Ori and Soen suggest that one could perhaps produce causality
violation without WEC violation, as follows: Let $U$ be the subset of
the Ori-Soen spacetime which does not violate WEC and which is not in
the future of any WEC-violating region.  Now imagine that we specify
the conditions on $U$, which we are presumably free to do, and then
let the rest of the spacetime evolve according to some WEC-obeying
equation of state.  If the resulting spacetime is continuous,
causality must still be violated, because the closed null geodesic $N$
lies on the boundary of $U$.  Thus while the Ori-Soen spacetime does
not violate the Tipler and Hawking theorems, it does seem to open a
possibility that causality violation could nevertheless be produced
using normal matter.  (See \cite{Krasnikov:1997ab} for another such
possibility.)

In this paper, I will study a simplified model with the same
properties as the Ori-Soen spacetime.  I will analyze its causal
structure and show how the unusual situation Ori and Soen describe can
arise.  I will then argue that the procedure described above may not
in fact result in a continuous spacetime at the place where the closed
null curve $N$ would otherwise appear.

\section{Simplified spacetime}

The spacetime of \cite{Ori1,Ori2,Ori3} has causality violation in a
toroidal region.  We will simplify the situation as follows: first we
will take the ``cylindrical'' metric of \cite{Ori2} in which the torus
has been straightened out.  We will take the $z$ axis to lie on the
axis of the cylinder.  This spacetime does not have CTC's, but we can
re-introduce them by making the $z$ direction periodic.  We will then
eliminate the azimuthal direction to produce a 2+1-dimensional
spacetime with CTC's lying in a strip.  We will use the metric
\bea\label{eqn:metric}
ds^2 &=& f (t)\left[-(dt-h (x)t\,dz)^2+dz^2+(dx-h (x)bx\,
dz)^2\right]\nonumber\\
&=& f (t)\big[-dt^2+2h (x) t\, dt\, dz\nonumber\\
&&\quad +(1-h (x)^2t^2+h (x)^2b^2x^2) dz^2\\
&&\quad -2h (x) bx\, dz\, dx+dx^2\big]\nonumber
\eea
with $t > 0$, $z$ periodic, and $f(t) > 0$.
The function $h(x)$ is a window function \cite{Ori1,Ori2,Ori3},
with a parameter $d$,
\be\label{eqn:h}
h (x)=\cases{[1-(x/d)^4]^3 & for $x < d$\cr
0 & for $x > d$.}
\ee
Outside the strip $x\in[-d, d]$, the
spacetime is conformal to Minkowski space.  The function $f(t)$
provides a conformal factor that does not affect the causal structure
but allows energy conditions to be satisfied.  We will take $f(1) = 1$
for simplicity, and adjust $f'(1)$ and $f''(1)$ as required.  Since
$-\det g = f (t)^3 > 0$ everywhere, the metric is never singular.

Since the $z$ coordinate is periodic, a path moving purely in
the $z$ direction is closed.  For points where $1 + h(x)^2[b^2x^2 - t^2]=
0$ this path is null.  At $x = 0$, $t = 1$ it is a closed null
geodesic which we call $N$, following \cite{Ori1,Ori2,Ori3}.
For larger $t$, such paths are closed timelike curves.

However, the intrinsic metric of a surface given by $t = t_0 < 1 $ is
\be
dr^2 = f(t)[(1-h^2t_0^2) dz^2+(dx-h (x) bx\, dz)^2]\,,
\ee
which is positive definite.  Thus the space with $t < 1$ can be
foliated in constant-time surfaces and so has normal causality.

To study the causal structure, we first note that nothing depends on
the $z$ coordinate.  We will be concerned only with the set of motions
in $t$ and $x$ that are possible within the light cone, when any motion in
$z$ is permitted.  Thus we project the light cone into the $t$-$x$
plane.  If the $z$ direction is timelike, then motion in any direction
in the $t$-$x$ plane is possible.  If not, there will be maximum and
minimum values of $dt/dx$ for a causal curve, which are given by
\be\label{eqn:dtdx}
{dt\over dx}\bigg|_{\stackrel{\scriptstyle\text{max}}{\text{min}}}
={h^2bxt\pm\sqrt{1+h^2b^2x^2-h^2t^2}\over{1+h^2b^2x^2}}\,.
\ee
We will also define a function $t_n (x) =\sqrt{h^{-2} +b^2x^2}$ which
gives the boundary of the region where the $z$ direction is timelike.

Now we can find the Cauchy horizon.  Points with $t\ge t_n$ lie on
closed causal curves with $t$ and $x$ constant, so any such point is
in the causality violating region.  It is clear that every point of
the spacetime has causality violating points in its future, so a
point is on a closed causal curve exactly if there is a point with
$t\ge t_n$ in its past, and the Cauchy horizon is just the boundary of
the causality violating region.

Now we integrate Eq.\ (\ref{eqn:dtdx}) to get the Cauchy horizon.
There are two different situations depending on the magnitude of
$dt/dx$.  As $t\rightarrow t_n$, the projected light cones open out so
that maximum and minimum values of $dt/dx$ become the same.  If this
value is smaller than $dt_n/dx$, then null rays leave the region where
the $z$ direction is timelike and go outward, whereas if it is larger,
then null rays leave the region where the $z$ direction is timelike
and go inward.

\begin{figure}
\begin{center}
\leavevmode\epsfbox{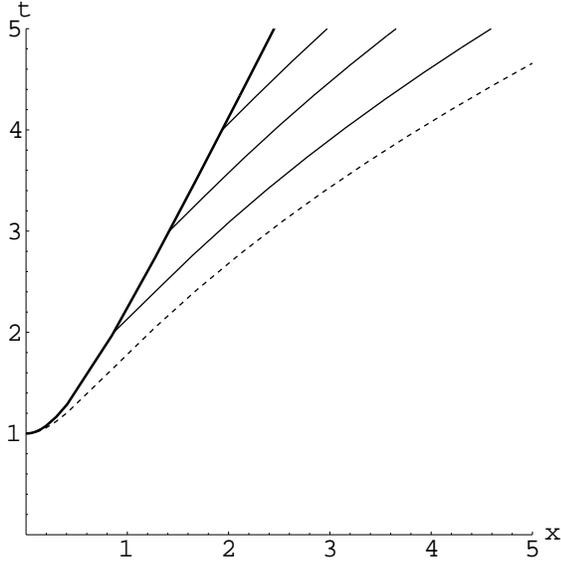}
\end{center}
\caption{Causal structure of the spacetime with $h = 1$ and $b > 1$.
In the region above the thick line, motion in the $z$ direction is
timelike.  The thin lines are null curves leaving this region.  The
dashed line is the Cauchy horizon.} 
\label{fig:bbig}
\end{figure}

This behavior can be understood by considering the case where $h (x) =
1$ everywhere, so that the entire spacetime, rather than just a strip,
is modified.  In this case we have $t_n (x) =\sqrt{1+b^2x^2}$, and at
$t = t_n (x)$,
\be
{dt\over dx}\bigg|_{\stackrel{\scriptstyle\text{max}}{\text{min}}}
={bxt\over{1+b^2x^2}}={bx\over\sqrt{1+b^2x^2}}
\ee
whereas
\be
{dt_n\over dx} ={b^2x\over\sqrt{1+b^2x^2}}\,.
\ee
Thus if $b > 1$ we have $dt/dx < dt_n/dx$, which leads to the
situation shown in Fig.\ \ref{fig:bbig}.
Future-directed null curves
cross $t = t_n$ in the direction of increasing $| x |$.  The Cauchy
horizon is the future-directed null curve from $N$ with $dt/dx$
minimal.  In this case, the Cauchy horizon is compactly generated: it
arises at the curve $N$.

\begin{figure}
\begin{center}
\leavevmode\epsfbox{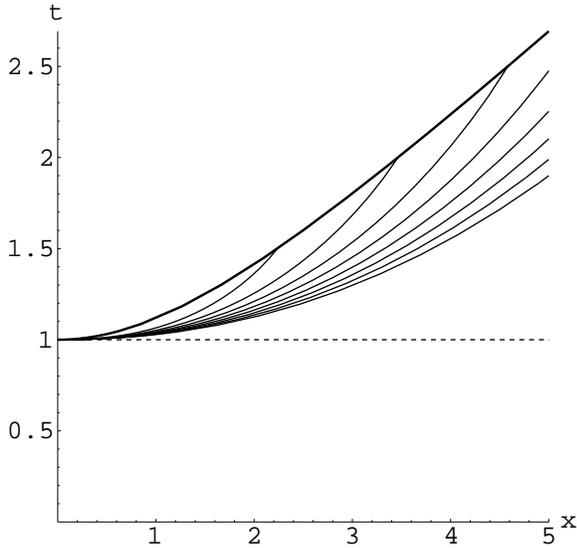}
\end{center}
\caption{Causal structure of the spacetime with $h = 1$ and $b < 1$.
In the region above the thick line, motion in the $z$ direction is
timelike.  The thin lines are null curves leaving this region and
going inward to $x = 0$.  The dashed line is the Cauchy horizon.} 
\label{fig:bsmall}
\end{figure}

In contrast, if $b < 1$, we have $dt/dx > dt_n/dx$, which leads to the
situation shown in Fig.\ \ref{fig:bsmall}.
Null curves cross $t =
t_n$ in the direction of decreasing $|x|$.  Any point with $t > 1$ is
reachable by such a curve, so the Cauchy horizon is just the surface
$t = 1$.  It comes in from infinity and thus is not compactly
generated.

It is straightforward to compute the stress-energy tensor for the
metric of Eq.\ (\ref{eqn:metric}) with $h = 1$.  First we look at
$T_{\mu\nu}$ at $t = 1, x = 0$ projected on the tangent vector to $N$
given by $V^z = 1$, $V^t = V^x = 0$,
\be
T_{\mu\nu} V^\mu V^\nu = T_{zz} ={1\over 8\pi G} b (1-b)\,.
\ee
If $b > 1$, $T_{zz} < 0$, so the weak energy condition cannot be obeyed, as one would expect since the Cauchy horizon is compactly generated in this case.

If $b < 1$, $T_{zz} > 0$, but we need also to look at other
projections of $T^{\mu\nu}$.  The matrix of mixed-index components
${T^\mu}_\nu$ at $t = 1$ is
\be
{1\over 32\pi G}
\left(\begin{array}{ccc}
4b^2+2f'(1+b) & 4b (1-b) & 0\\
3f'^2-2f'' & 4b+2f' (1+b) & 0\\
bx (3f'^2-2f'') & -2b (1-b) x (f' +2) & 4+4f'
\end{array}\right)\,.
\ee
The eigenvalues of ${T^\mu}_\nu$ are $-\rho$, $p_1$, and $p_2$, where
$\rho$ is the energy density and $p_i$ are the principal pressures.
In terms of the density and pressures, the weak energy condition can
be written $\rho\ge 0$ and $\rho +p_i\ge 0$.  We can write also the
dominant energy condition, $\rho\ge 0$ and $| p_i |\le\rho$, and the
strong energy condition, $\rho +p_i\ge 0$ and $\rho +\sum p_i\ge
0$.  We can choose parameter values so that all these conditions will be
satisfied.  For example,
\be\label{eqn:parameters}
b = 0.02, f' = -1, f'' = 0.6
\ee
give $\rho = 0.59$, $p_1 = 0$, and $p_2 = -0.41$.

\begin{figure}
\begin{center}
\leavevmode\epsfbox{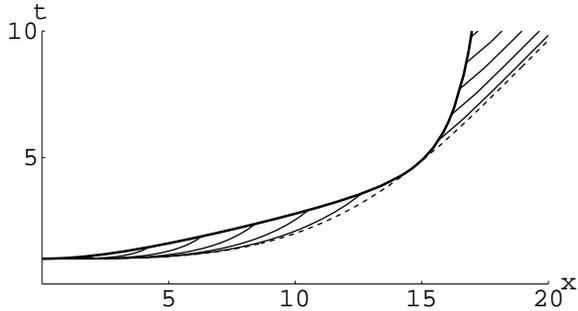}
\end{center}
\caption{Causal structure of the full spacetime.  In the region above
the thick line, motion in the $z$ direction is timelike.  The thin
lines are null curves leaving this region.  The dashed line is the
Cauchy horizon, which is tangent to the thick line at $N'$.  To the
left of $N'$, the future-directed null curves go inward and backward
in $t$; to the right, they go outward and forward in $t$.}
\label{fig:bh}
\end{figure}

Now we include the full form of $h(x)$ from Eq.\ (\ref{eqn:h}).
At $t = t_n (x)$, we have
\bml\label{eqn:dtdxh}\bea
{dt\over dx}\bigg|_{\stackrel{\scriptstyle\text{max}}{\text{min}}}
&=&{hbx\over\sqrt{1+h^2b^2x^2}}\\
{dt_n\over dx} &=&{-h^{-2}h' +hb^2x\over\sqrt{1+h^2b^2x^2}}\,.
\elea
Null curves leaving the region where the $z$ direction is timelike
go inward if
\be\label{eqn:criterion}
1-b >{-h'\over h^3bx}\,.
\ee
As $x\rightarrow 0$, $h'\sim x^3$, so the right hand side of Eq.\
(\ref{eqn:criterion}) vanishes, and the condition is satisfied if $b <
1$.  However, as $x\rightarrow d$, $h'/h\rightarrow -\infty$, so the
condition is never satisfied.  Thus null curves leaving the region
where the $z$ direction is timelike go inward for small values of $x$,
but outward for large values of $x$.  There is some intermediate point
where $-h'/(h^3bx) = 1-b$, from which the two families of null curves
diverge.  Following \cite{Ori2} we will call this point (really a
curve with fixed $x$ and $t$) $N'$.  This situation is shown in Fig.\
\ref{fig:bh}.

\begin{figure}
\begin{center}
\leavevmode\epsfbox{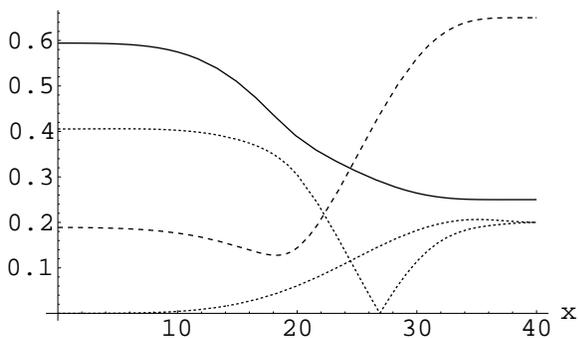}
\end{center}
\caption{Density and pressures for the parameter values of Eq.\
(\ref{eqn:parameters}) and $d = 40$.  The solid line is $\rho$, the
dotted lines are $| p_i |$, and the dashed line is $\rho +p_1+p_2$.}
\label{fig:eigenvalues}
\end{figure}

In this case, the Cauchy horizon arises from $N'$ and spreads out in
both directions.  The generators going inward terminate at $x = 0$,
but the generators going outward continue to infinity.  By the
theorems of Hawking and Tipler, the weak energy condition must be
violated at $N'$, but it is possible to obey the weak, strong, and
dominant energy conditions in a neighborhood of the slice $t = 1$.  To
see this, we can compute the eigenvalues of the full ${T^\mu}_\nu$,
e.g., using Mathematica.  With the parameter values of Eq.\
(\ref{eqn:parameters}) and $d = 40$, we find the results shown in
Fig.\ \ref{fig:eigenvalues}.
We see that for all values of $x$ we have $\rho > 0$, and
$\rho > | p_i |$, so the dominant and weak energy conditions are
satisfied.  We also see that $\rho +p_1+p_2 > 0$, so the strong energy
condition is satisfied.  Since all these conditions are satisfied
strongly, i.e., with $>$ instead of $\ge$, they must be satisfied over
some finite range of $t$ near $1$.

\section{Discussion}

From the point of view of the Tipler/Hawking theorems, the Ori-Soen
spacetime is perfectly normal.  It contains a region of closed
timelike curves with a compactly-generated Cauchy horizon, and it
violates the weak energy condition where the horizon arises.  However,
there still remains the original question of whether one can use this
spacetime to generate causality violation with normal matter.

\begin{figure}
\begin{center}
\leavevmode\epsfbox{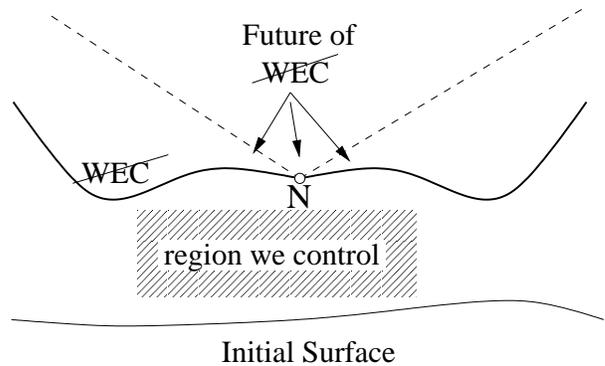}
\end{center}
\caption{The remaining spacetime after the WEC violating region and
its future have been removed.  We can control all the space below and
including the thick line, except for the point $N$.}
\label{fig:last}
\end{figure}

To make such an attempt, one removes from the spacetime all the areas where
the weak energy condition is violated, and all areas that could be
influenced by the WEC-violating regions.  Thus one must remove at
least a region around the point $N'$, and its causal future.  This
future contains $N$, but does not include any point with $x = 0$
and $t < 1$.  We thus have the situation shown in Fig.\
\ref{fig:last}.
As discussed in \cite{Ori1,Ori2}, the causality violation occurs in
the boundary of the region that we can control.  In this case, we have
just the single closed null curve $N$ that has this property; the
entire rest of the causality-violating region has been removed. The
system has no chronology violation, since $N$ is null.  But it does
have {\em causality} violation at $N$, which cannot be avoided unless
there is a metric discontinuity there.

\begin{figure}
\begin{center}
\leavevmode\epsfbox{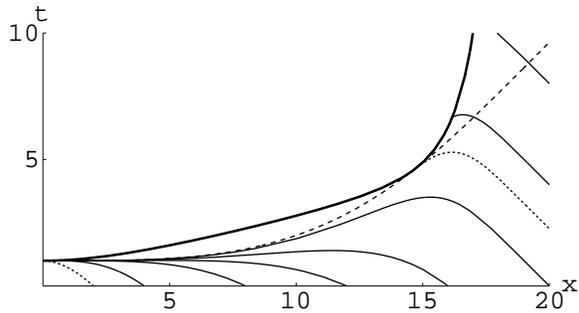}
\end{center}
\caption{Causal structure preceding the region of causality violation.
Thin lines are the most left-going null paths.  The dotted line near
the origin is the boundary of the past of points immediately before
$N$, and the dotted line at the right is the boundary of the past of
$N$ in the unmodified spacetime.  Removing some space around $N$ might
reduce the volume of the past of $N$, but there will still be a
discontinuity.}
\label{fig:causal}
\end{figure}

However, there is some reason to believe that such a discontinuity
might be expected.  As shown in Fig.\ \ref{fig:causal},
the past volume of $N$ is much larger than the past volume of points
on the $t$ axis immediately before $N$.  Past-volume discontinuities
always exist when there is a CTC-containing region, because every
point in such region is in the past of every other point.  However,
the present situation is different in that the discontinuity has
appeared already on the supposed boundary of the CTC-containing
region, and does not depend on the interior.  As noted in
\cite{Ori2}, $N$ is also a curve where many null hypersurfaces
intersect.

It seems reasonable to imagine that, in a realistic field theory, the
past-volume discontinuity would lead to a piling up of modes at $N$,
and consequently a discontinuity in the metric, which would prevent
the curve $N$ from being causal.  Similar dynamics have been discussed
in \cite{cpc}, although there is some question about how general such
dynamics are \cite{LiGott}.  Another way to make this
argument\footnote{This idea is due to L. H. Ford} is to note that in a
continuous spacetime, it should be possible to make small
perturbations on the initial surface to produce another similar
spacetime.  But here, because of the divergence in past volume, small
perturbations would not remain small at the point $N$.

The past-volume discontinuity is not accidental, but is required for
causality violation to take place on the boundary of the controlled
region.  If the past volumes were continuous, then all points near the
Cauchy horizon would be in the future of points near the fountain.
When one excised the WEC-violating region and its future, one would
remove a neighborhood around the entire Cauchy horizon, and so
causality violation could not take place on the boundary of the
controlled region.

In spite of the above arguments, I cannot rule out the alternative
possibility that a naked singularity could form in place of the region
surrounding $N'$ where the weak energy condition is violated.  The
generators of the Cauchy horizon could then arise from this
singularity, and the theorems of Hawking and Tipler would be
satisfied, because the singularity renders the region non-compact.  In
such a case, we would have succeeded in creating causality violation
using normal matter, as described in the introduction.  Although the
details of the causality violation could not be predicted because of
the singularity, we could predict that causality violation will occur,
as long as there is no metric discontinuity at $N$ as above.

\section*{acknowledgments}

I would like to thank Arvind Borde, Larry Ford, Alan Guth, Allen
Everett, and Tom Roman for helpful conversations.  This work was
supported in part by funding provided by the National Science
Foundation.


\begin{thebibliography}{1}

\bibitem{tip76}
F.~J. Tipler, Phys. Rev. Lett. {\bf 37},  879  (1976).

\bibitem{tip77}
F.~J. Tipler, Ann. Phys. {\bf 108},  1  (1977).

\bibitem{cpc}
S.~W. Hawking, Phys. Rev. D {\bf 46},  603  (1992).

\bibitem{Ha&El}
S.~W. Hawking and G.~F.~R. Ellis, {\em The Large Scale Structure of Space-time}
  (Cambridge University Press, London, 1973).

\bibitem{Ori1}
A. Ori, Phys. Rev. Lett. {\bf 71},  2517  (1993).

\bibitem{Ori2}
A. Ori and Y. Soen, Phys. Rev. D {\bf 49},  3990  (1994).

\bibitem{Ori3}
A. Ori and Y. Soen, Phys. Rev. D {\bf 54},  4858  (1996).

\bibitem{Krasnikov:1997ab}
S.~V. Krasnikov, Class. Quant. Grav. {\bf 15},  997  (1998).

\bibitem{LiGott}
L.-X. Li and J.~R. Gott, III, Phys. Rev. Lett. {\bf 80},  2980  (1998).

\end{thebibliography}
\end{document}